\documentclass[twoside,fleqn]{article}
\usepackage{espcrc2,epsf}

\newenvironment{ul}{\begin{list}{$\bullet$}
 {\settowidth{\labelwidth}{$\bullet$}\leftmargin\labelwidth
 \addtolength{\leftmargin}{0.5em}%
 \itemsep0pt\parsep0pt\topsep0.8ex}}{\end{list}}

\def\btorho{\bar B^0\to\rho^+ l^- \bar\nu_l}
\def\btokstargamma{B\to K^*\gamma}
\def\btopi{\bar B^0\to\pi^+ l^- \bar\nu_l}
\def\vub{|V_{ub}|}
\def\qsqmax{q^2_{\rm max}}
\def\gev{\,\mathrm{Ge\kern-0.1em V}}
\def\dof{\mathrm{dof}}
\def\erparen#1#2{\relax\ifmmode{}(^{#1}_{#2})\else$(^{#1}_{#2})$\fi}
\def~{\ifmmode\phantom{0}\else\penalty10000\ \fi}

\newdimen\unit
\def\point#1 #2 #3{\vbox to0pt{\kern-#2\unit
  \hbox{\kern#1\unit$#3$}\vss}\nointerlineskip}

\title{Lattice-Constrained Parametrizations of Form Factors for
Semileptonic and Rare Radiative $B$ Decays}

\author{UKQCD Collaboration:
L Del Debbio\address{Centre de Physique Th\'eorique,
 CNRS Luminy, Case 907, F-13288 Marseille Cedex 9,
 France (Unit\'e Propre de Recherche 7061)},
J M Flynn\address{Department of Physics \& Astronomy, University of
 Southampton, Southampton SO17 1BJ, UK} (presenter),
L Lellouch$^{\mathrm{a}}$
and
J Nieves\address{Departamento de Fisica Moderna, Avenida
 Fuentenueva, 18071 Granada, Spain}} 
       
\begin{document}

\begin{abstract}
We describe the form factors for $\btorho$ and $\btokstargamma$ decays
with just two parameters and the two form factors for $\btopi$ with
three parameters. The parametrizations are constrained by lattice
results and are consistent with heavy quark symmetry, kinematic
constraints and light cone sum rule scaling relations.
\end{abstract}

\maketitle

We obtain a simple yet phenomenologically useful description of
the form factors for semileptonic and rare radiative heavy-to-light
meson decays for all $q^2$, the squared momentum transfer to the
leptons or photon. Lattice calculations determine the form factors
over a limited region at high $q^2$. We use model input to extend
the results to $q^2{=}0$, seeking consistency with:
\begin{ul}
\item kinematic constraints: $F_1(0) = F_0(0)$ and $T_1(0) = i T_2(0)$
\item heavy quark symmetry (HQS)
\item light cone sum rule (LCSR) scaling relations: all form factors
scale like $M^{-3/2}$ as $M {\to}\infty$ at $q^2{=}0$, where $M$ is the
heavy meson mass
\item dispersive constraints
\end{ul}
The normalisation is fixed using lattice results.  The outcome is a
two parameter fit for $\btorho$ or $\btokstargamma$ and a three
parameter fit for $\btopi$. More details can be found
in~\cite{ukqcd:hlfits}.

The leading order HQS analysis shows that heavy-to-light $P{\to} P$
decay form factors are determined by two universal (``Isgur-Wise'')
functions, while $P{\to} V$ decays are governed by four more such
functions ($P$ and $V$ denote pseudoscalar and vector mesons
respectively). We adopt a model of Stech~\cite{stech1} which keeps
just one universal function for $P{\to} P$ and one more for $P{\to}
V$.

Lattice simulation details can be found
in~\cite{ukqcd:hlff,ukqcd:btorho} with details of the chiral
extrapolation for $\btopi$ in~\cite{lpl:btopi-bounds}.  All form
factors are calculated for four values of the heavy quark mass around
the charm mass and for a variety of $q^2$.  In our previous work
\cite{ukqcd:hlff,ukqcd:btorho}, the form factors were extrapolated at
fixed four-velocity recoil, $\omega = v \mathord\cdot
(p_{P,V}/m_{P,V})$, near the zero recoil point $\omega=1$, using the
heavy-quark scaling relations:
\[
f\Theta M^{n_f/2}=
 \gamma_f\left(1+\frac{\delta_f}{M}+\frac{\epsilon_f}{M^2}+\cdots\right)
\]
where $n_f=-1,1,-1,-1,1,-1,-1,1$ for $f=F_1,F_0,A_0,V,A_1,A_2,T_1,T_2$
and $\gamma_f$, $\delta_f$ and $\epsilon_f$ are fit parameters.
$\Theta$ comes from leading logarithmic matching and is chosen to be 1
at the $B$ mass. This procedure neglects the fact that for
$M{\to}\infty$, HQS predicts $A_1=2iT_2$ and $V=2T_1$ at fixed
$\omega$ not too far from $\qsqmax$. We enforce this condition by
performing a combined fit, at fixed $\omega$, of the pairs $(A_1,T_2)$
and $(V,T_1)$ imposing the constraints: $\gamma_{A_1}=2i\gamma_{T_2}$
and $\gamma_V=2\gamma_{T_1}$.  This guarantees that the extrapolated
form factors are consistent with HQS in the infinite mass limit and
reduces statistical errors by decreasing the number of parameters.

\section{$\btorho$ AND $\btokstargamma$ DECAYS}

\begin{figure}
\unit=\hsize
\hbox to\hsize{\hss
\vbox{\offinterlineskip
\epsfxsize=\unit\epsffile{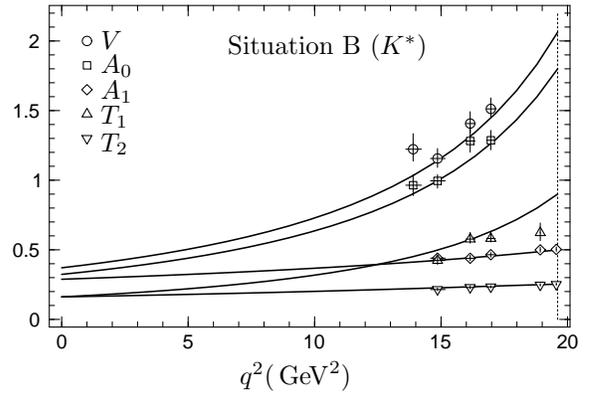}
\point 0.38 0.58 {{\rm Situation\ B\ }(K^*)}
\point 0.155 0.584 V
\point 0.155 0.542 {A_0}
\point 0.155 0.497 {A_1}
\point 0.155 0.452 {T_1}
\point 0.155 0.404 {T_2}
%\kern0.5em
\hbox to\unit{\hfill$q^2 (\gev^2)$\hfill}
}\hss}\kern-2em
\caption[]{Fit to the lattice predictions for $A_0$, $A_1$, $V$, $T_1$
and $T_2$ for a $K^*$ meson final state (Situation B) assuming a pole
form for $A_1$. The dashed vertical line indicates $\qsqmax$.
\label{fig:bsg}}
\end{figure}

\begin{table*}
\caption[]{Form factor results for $\btorho$ and $\btokstargamma$. For
$\btorho$ the fit parameters are: $A_1(0) = 0.27\erparen{5}{4}$, $M_1
= 7.0\erparen{12}{~6}\gev$, $\chi^2/\dof = 24/20$. For
$\btokstargamma$: $A_1(0) = 0.29\erparen{4}{3}$, $M_1 =
6.8\erparen{7}{4}\gev$, $\chi^2/\dof = 27/20$.}
\label{tab:vecfits}
\begin{center}\renewcommand{\arraystretch}{1.2}
\begin{tabular}{lllllll}
\hline
$q^2$ & $A_1$ & $A_2$ & $A_0$ & $V$ & $T_1$ & $T_2$ \\ \hline
$0$ & $0.27\erparen{5}{4}$ & $0.26\erparen{5}{3}$  &
 $0.30\erparen{6}{4}$ & $0.35\erparen{6}{5}$ &
 $0.16\erparen{2}{1}$ \\
$\qsqmax$ & $0.46\erparen{2}{1} $ & $0.88\erparen{5}{3}$ & 
 $1.80\erparen{9}{5}$ & $2.07\erparen{11}{~6}$ &
 $0.90\erparen{5}{4}$ & $0.25\erparen{1}{1}$ \\
\hline
\end{tabular}
\end{center}
\end{table*}
We use the freedom to adjust quark masses in lattice calculations and
consider two situations for the light quark $q$ into which the $b$
decays:
\settowidth{\labelwidth}{{\bf B}}\leftmargini\labelwidth
 \addtolength{\leftmargini}{0.5em}%
\begin{itemize}\itemsep0pt
\item[{\bf A}] $q{=}u$: matrix elements of $\bar
u\,\sigma^{\mu\nu}(1{+}\gamma^5)b$ are unphysical but constrain $\btorho$.
\item[{\bf B}] $q{=}s$: $\bar s\gamma^{\,\mu}(1-\gamma^{\,5}) b$
is unphysical but constrains $\btokstargamma$.
\end{itemize}
We complete the parametrization by specifying one of the form
factors. To meet all our requirements, including the LCSR scaling
condition at $q^2=0$, we choose
\begin{equation}
A_1(q^2) = 
\frac{A_1(0)}{1-q^2/M_1^2}
\label{eq:a1pole}
\end{equation}
with free parameters $A_1(0)$ and $M_1$. This allows $A_1$, $A_2$ and
$T_2$, which receive contributions from $1^+$ resonances, to diverge at
larger $q^2$ than the more singular $V$, $A_0$ and $T_1$. We also
tried other parametrizations but all results below will use $A_1(q^2)$
in eq.~(\ref{eq:a1pole}). Figure~\ref{fig:bsg} shows the fit for a
final state with the mass of the $K^*$, and Table~\ref{tab:vecfits}
gives results for the form factors.

\section{$\btopi$ DECAYS}

Stech's model makes $F_0(\qsqmax)$ vanish in the chiral limit,
contradicting our results and made unlikely by unitarity
bounds~\cite{lpl:btopi-bounds}. Furthermore, the $B^*$ which
contributes a pole very close to $\qsqmax$ in $F_1$, induces the same
singularity in $F_0$ in the model. This provokes a much stronger $q^2$
dependence for $F_0$ than seen in the lattice results or induced by
the nearest $0^+$ resonance.  Therefore we restrict to polar-type
$q^2$-dependences, consistent with the kinematical constraint,
$F_1(0)=F_0(0)$, HQS and unitarity bounds. Our preferred model,
consistent with LCSR scaling relations at $q^2=0$, is
\[
F_1(q^2) = \frac{F(0)}{(1{-}q^2/m_1^2)^{2}}, \quad
F_0(q^2) = \frac{F(0)}{(1{-}q^2/m_0^2)}.
\]
The result of the fit is: $F(0) = 0.27(11)$, $m_1 = 5.79(58)\gev$ and
$m_0 = 6.1(15)\gev$ with $\chi^2/\dof = 0.1/3$.  All results below
will be quoted using this pole/dipole model.

\section{PHENOMENOLOGY}

\begin{table*}
\caption[]{Form factor values at $q^2 =0$ with $B\to \pi,\rho$
semileptonic decay rates and ratios from this calculation and from
light cone sum rules (LCSR). Decay rates are given in units of $\vub^2
\,\mathrm{ps}^{-1}$. $\Gamma_{\rho/\pi} \equiv
\Gamma(\btorho)/\Gamma(\btopi)$ and $\Gamma_{L/T}$ denotes the ratio
of rates to longitudinally and transversely polarised rho mesons in
$\btorho$. $l$ denotes a massless lepton.}
\label{tab:lcsrcf}
\begin{center}
\def\arraystretch{1.16}\tabcolsep0.75\tabcolsep
\begin{tabular}{@{}llllllllllll@{}}
\hline
 & $F_1(0)$ & $A_1(0)$ & $A_2(0)$ & $V(0)$ & $T_1(0)$ &
 $\Gamma_{\pi l\bar\nu}$ & $\Gamma_{\rho l\bar\nu}$ &
 $\Gamma_{\rho/\pi}$ & $\Gamma_{L/T}$ &
 $\Gamma_{\pi\tau\bar\nu_\tau}$ & $\Gamma_{\rho\tau\bar\nu_\tau}$ \\
\hline
 & $0.27(11)$ & $0.27\erparen{5}{4}$ & $0.26\erparen{5}{3}$ &
 $0.35\erparen{6}{5}$ & $0.16\erparen{2}{1}$ &
 $8.5\erparen{33}{14}$ & $16.5\erparen{35}{23}$ & $1.9\erparen97$ &
 $0.80\erparen43$ & $5.8 \erparen{18}{~4}$ & $8.8 \erparen{14}{~9}$
 \\[0.8ex]
\multicolumn{2}{@{}l}{LCSR} \\
%LCSR
\cite{ballbraun:lcsr} & & $0.27(5)$ & $0.28(5)$ & $0.35(7)$ & &
  & $13.5(40)$ & $1.7(5)$ & $0.52(8)$\\
%LCSR
\cite{alibraunsimma:lcsr} & & $0.24(4)$ & & $0.28(6)$ & $0.16(3)$ \\
%LCSR
\cite{bbkr:lcsr} & \multicolumn{2}{l}{$0.24$--$0.29$} & & & & $8.7$ \\
%LCSR
\cite{becirevic} & & & & & 0.15(3) \\
\hline
\end{tabular}
\end{center}
\end{table*}
\begin{figure*}
\unit=\hsize
\hbox to\hsize{\hfill
\vbox{\offinterlineskip
\epsfxsize=\unit\epsffile{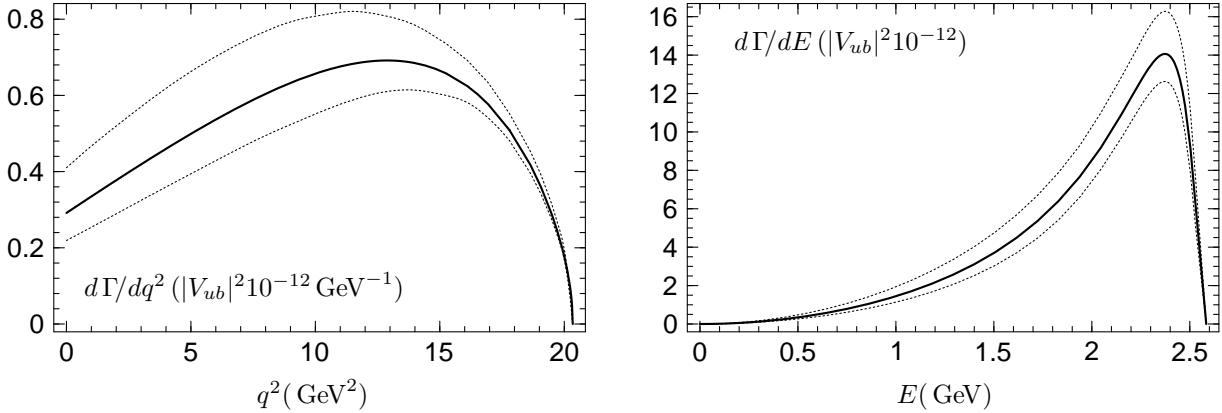}
\point 0.06 0.08 {d\,\Gamma\!/dq^2 \, (\vub^2 10^{-12}\gev^{-1})}
\point 0.6 0.283 {d\,\Gamma\!/dE \, (\vub^2 10^{-12})}
\kern0.4em
\hbox to\unit{\hbox to0.5\unit{\hfill$q^2 (\gev^2)$\hfill}
\hfill\hbox to0.45\unit{\hfill$E (\gev)$\hfill}}
}\hfill}
\kern-2em
\caption[]{Differential decay spectra for $\btorho$ for massless
leptons: (a) $d\,\Gamma/dq^2$ in units of $10^{-12} \vub^2 \gev^{-1}$,
(b) $d\,\Gamma/dE$ in units of $10^{-12} \vub^2$. The dashed
lines show the envelope of the 68\% bootstrap errors computed
separately for each value of $q^2$ or $E$
respectively.\label{fig:btorho-spectra}}
\end{figure*}
Using our fits we can calculate total rates and differential decay
spectra in $q^2$ and lepton energy $E$ for the decays $\btorho$
(Figure~\ref{fig:btorho-spectra}) and $\btopi$. In
Table~\ref{tab:lcsrcf} we give our results, illustrating the good
agreement of our form factor values at $q^2{=}0$ with LCSR
calculations, and compare rates and ratios for semileptonic
decays. For $\btokstargamma$ we evaluate the ratio $R_{K^*} =
\Gamma(B\to K^*\gamma)/\Gamma(b\to s\gamma) = 16\erparen43\%$, to be
compared with $(18\pm 7)\%$ from experiment~\cite{CLEO}.

\end{document}